\documentclass[twocolumn,showpacs,preprintnumbers,amsmath,amssymb]{revtex4}

\usepackage{graphicx}
\usepackage{bm}

\begin{document}


\title{Rashba type spin-orbit splitting of quantum well states in ultrathin Pb films}

\author{Hugo Dil$^{1,2}$}
\author{Fabian Meier$^{1,2}$}
\author{Jorge Lobo-Checa$^{3}$}
\author{Luc Patthey$^{1}$}
\author{Gustav Bihlmayer$^{4}$}
\author{J\"urg Osterwalder$^{1}$}
\affiliation{$^{1}$Physik-Institut, Universit\"at Z\"urich, Winterthurerstrasse 190, CH-8057 Z\"urich, Switzerland \\ $^{2}$ Swiss Light Source, Paul Scherrer Institut, CH-5232 Villigen, Switzerland \\ $^{3}$Departement Physik, Universit\"at Basel, Klingelbergstrasse 82, CH-4056 Basel, Switzerland\\$^{4}$ Institut f\"ur Festk\"orperforschung, Forschungszentrum J\"ulich, D-52425 J\"ulich, Germany}

\date{\today}

\begin{abstract}
A Rashba type spin-orbit splitting is found for quantum well states formed in ultrathin Pb films on Si(111). The resulting momentum splitting is comparable to what is found for semiconductor heterostructures. The splitting shows no coverage dependency and the sign of the spin polarization is reversed compared to Rashba splitting in the Au(111) surface state. We explain our results by competing effects at the two boundaries of the Pb layer.
\end{abstract}

\maketitle

The spin degeneracy in free electron like bands in solids can be lifted due to the breaking of the space inversion symmetry. For crystals which lack an inversion symmetry centre in their unit cell, this is referred to as the Dresselhaus effect\cite{Dresselhaus}. This symmetry can also be broken at an interface or surface and the resulting spin splitting of the bands is based on the Rashba-Bychkov effect\cite{Rashba} (henceforth Rashba effect). Using de Haas-van Alphen oscillations or photoluminescence, a Rashba splitting has been observed in the 2-dimensional electron gas in semiconductor heterojunctions\cite{Nitta}.\\
Using the more direct method of angle resolved photoemission (ARPES), a spin-orbit splitting of the $L$-gap surface state of Au(111)\cite{LaShell} and the surface states on W(110)\cite{Rotenberg} and bismuth\cite{Koroteev} have been reported among others. The spin splitting of these surface states has been confirmed by spin resolved ARPES (SARPES)\cite{Hochstrasser,Hoesch1,Hirahara1}. More recently a strong enhancement of the Rashba splitting has been reported for the Bi/Ag(111) surface alloy\cite{Ast}, which is largely attributed to a combination of in-plane and out-of-plane symmetry breaking\cite{Premper}.\\
Quantum well states (QWS) are standing electron waves that can form when the thickness of a metal layer becomes comparable to the electron coherence length\cite{Chiang}. The states are quantized in the direction perpendicular to the surface, but parallel to the surface the bands typically show a free electron like dispersion\cite{Dil1}. The confinement potential for QWS is asymmetric in the sense that on one side it is constituted by the metal-substrate interface and on the other side by the metal-vacuum interface. In many aspects QWS and surface states are thus very similar, and in some models surface states are regarded as zero thickness QWS\cite{Echenique}. This makes the fact that up to this moment to our best knowledge no Rashba type splitting of QWS has been reported\cite{Hirahara1, Hirahara2} most surprising. This absence has been attributed to the idea that the charge density is located too far from the surface or interface to experience the potential gradient\cite{Hirahara2} or to the interpretation that QWS are standing waves and should therefore show no Rashba splitting\cite{Petersen}.\\ 
It has been shown that QWS can become spin polarized through a hybridization with the spin split surface (or interface) states of the substrate\cite{Koitzsch,Shikin}. This effect diminishes with increasing layer thickness as the influence of the substrate on the electronic structure decreases, and can also be interpreted as a good example of how information about the interface can be transported to the surface by QWS. We will show here that in Pb a thickness independent Rashba type spin orbit splitting, similar to what is found in semiconductor heterostructures, is present in QWS, but that the energy separation between the bands is smaller than the intrinsic line width\cite{Reinert} and can therefore not be detected by spin integrated photoemission.\\
The experiments have been performed at T $<100$ K using the COPHEE spectrometer\cite{Hoesch2} at the surface and interface spectroscopy (SIS) beamline at the Swiss Light Source, using horizontally polarized light at 24 eV. This spectrometer is capable of recording the three spatial components of the spin polarization vector for any point in reciprocal space. In this way a spin resolved band structure can be obtained\cite{comment1}.\\
Figure 1(a) shows a spin integrated binding energy versus in-plane momentum plot for a 10 monolayer (ML) thick film of Pb on Si(111)($\sqrt{3}\times\sqrt{3}$)R30$^{\circ}$:Pb. (Henceforth referred to as Si(111)$\sqrt{3}$). The preparation and properties of this system have been described elsewhere\cite{Upton, Dil2}. Due to the layer-by-layer growth at low temperature the QWS binding energy position provides an intrinsic thickness calibration. In Figure 1(a), a single QWS can be identified with a binding energy of the $p_{z}$ derived bands of 0.18 eV at normal emission. From 0.5 \AA $^{-1}$ onwards the image is dominated by the strongly dispersing $p_{x,y}$ derived bands. Characteristic for this system is the high effective mass of the QWS around the centre of the surface Brillouin zone (SBZ) of up to 10 times the free electron mass\cite{Dil2}.\\
\begin{figure}[htb]
\begin{center}
\includegraphics[width=0.45\textwidth]{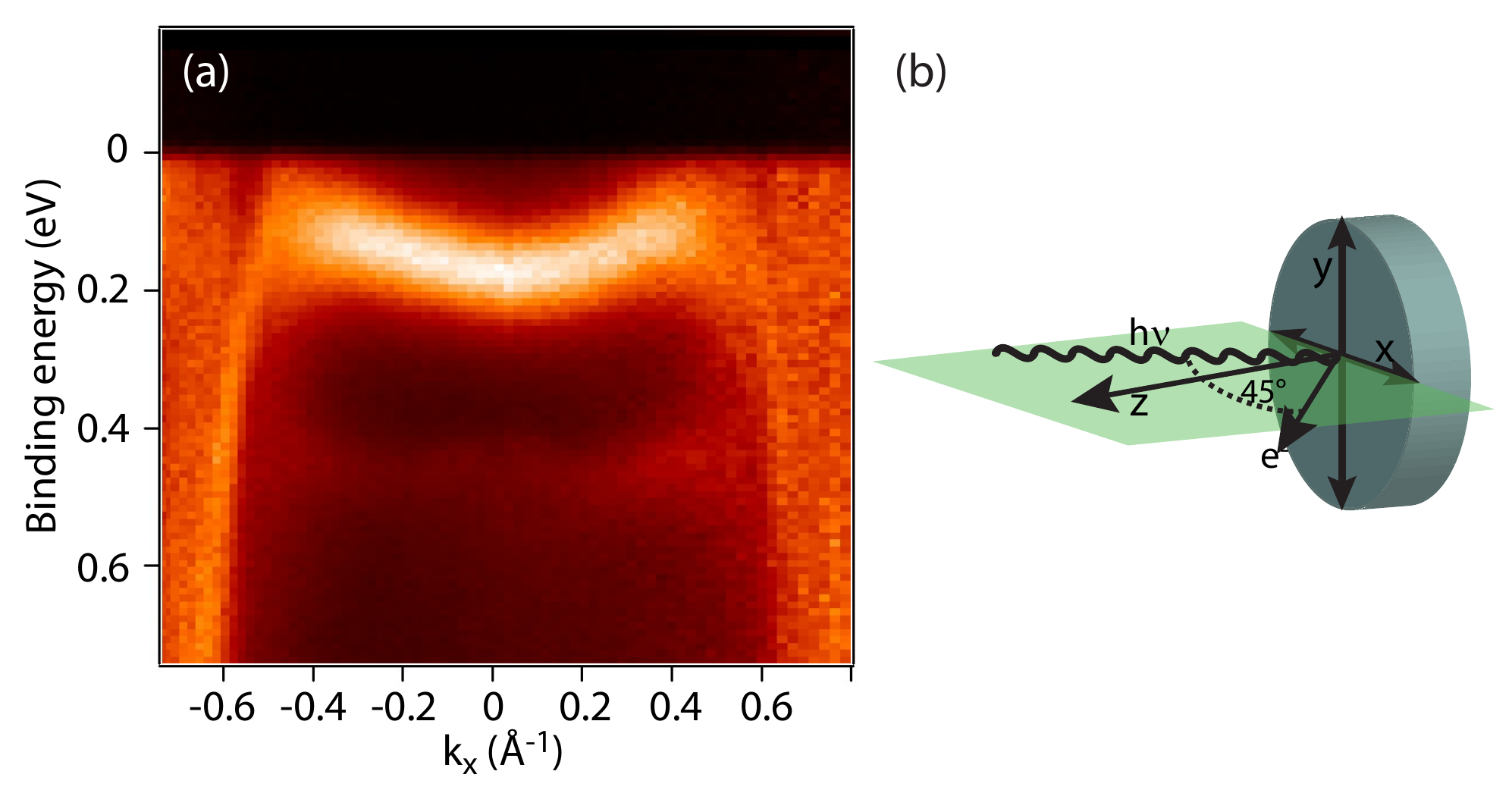} 
\caption{{(color online) (a) Measured spin integrated ARPES data for a 10 ML thick Pb film on Si(111)$\sqrt{3}$. (b) Measurement geometry and sample coordinates.}}
\label{Fig1}
\end{center}
\end{figure}
In the film plane the states can be considered as a two-dimensional nearly free electron gas for small $\bf{k_{||}}$ values. The spin dependent part of the Hamiltonian can therefore be reduced to $H_{R}=\alpha_{R}\bf{\sigma}\cdot(\bf{k_{||}}\times\bf{e_{z}})$ \cite{Petersen}, where $\bf{\sigma}$ is the vector of Pauli matrices and $\alpha_{R}$ is the Rashba parameter, which is within this model proportional to the potential gradient perpendicular to the surface, $\bf{e_{z}}$. This results in the formation of two free electron like parabolae that are shifted by $k_{0}$ from the centre of the surface Brillouin zone. The Rashba parameter can be expressed as $\alpha_{R}=\hbar^{2} k_{0}/m^{*}$(with $m^{*}$ the effective mass) and is thus solely based on parameters that are accessible in ARPES.\\
Figure 2(a) and (b) show spin polarization spectra for the $x$, $y$ and $z$ (out-of-plane) direction of the crystal (see drawing in Figure 1(b)), taken at $k_{y}=-0.08$ \AA $^{-1}$ and $k_{x} = 0$ for a 8 ML thick Pb film on Si(111)$\sqrt{3}$. In the $y$ and $z$ direction no signal can be discerned that is larger than the statistical error margins. The absence of any spin polarization in the out-of-plane direction is characteristic for all our measurements on this system, hence this component will not be discussed in the rest of this work. For the $x$-component a clear polarization signal is observed with an amplitude of approximately 10\%. Because $k_{x} = 0$, $k_{y}\neq0$ and $P_{y} = P_{z} = 0$ the spin quantization axis is oriented along the $x$ direction of the crystal for this point in reciprocal space, as illustrated in the schematic constant energy surface in Figure 2(d). A spin resolved spectrum can thus be calculated from $P_{x}$ according to $I_{up} = I_{tot}(1 + P_{x})/2$ and $I_{down} = I_{tot}(1 - P_{x})/2$. The resulting spectra are shown in Figure 2(c), where two distinct peaks can be discriminated with a separation of 12 meV.\\
\begin{figure}[htb]
\begin{center}
\includegraphics[width=0.45\textwidth]{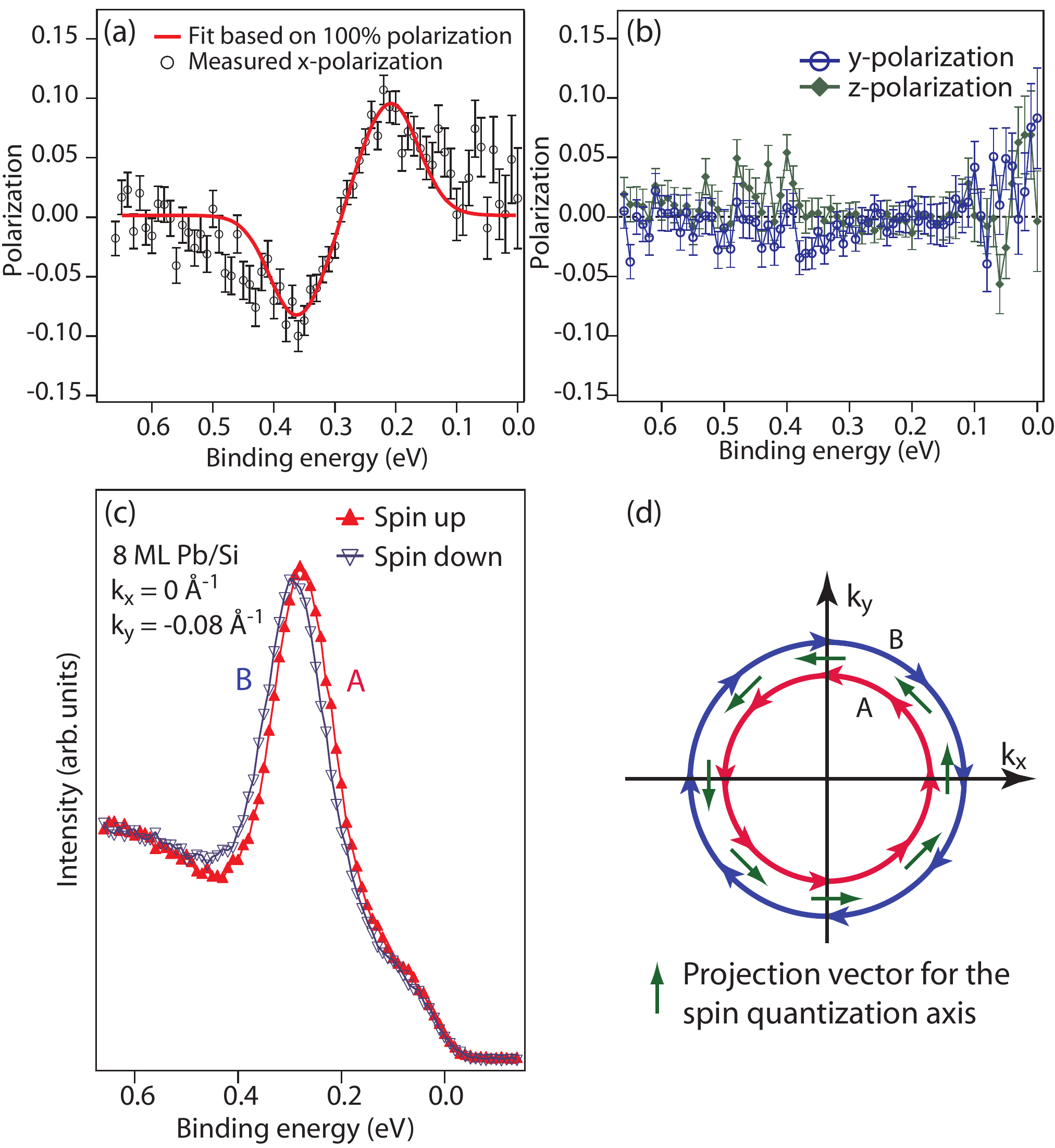} 
\caption{{(color online) SARPES data for an 8 ML thick Pb layer on Si(111)$\sqrt{3}$ at $k_{y}=-0.08$ \AA $^{-1}$ and $k_{x} = 0$. (a) Measured (open circles) and modeled (solid line) spin polarization in the $x$-direction of the sample. (b) Measured spin polarization along the $y$ (blue circles) and $z$ (green diamonds) direction of the sample. (c) Spin resolved spectra obtained from the spin polarization in the $x$-direction. (d) Schematic representation of a constant energy surface where the arrows of band A and B refer to the direction of the spin polarization axis.}}
\label{Fig2}
\end{center}
\end{figure}
Now the spin integrated spectrum can be fitted with two Gaussians, an unpolarized background and a Fermi function. In turn the spin polarization spectra can be fitted in a similar manner as described elsewhere\cite{Meier}. The best correspondence between model and data in Figure 2(a) is obtained when assuming a 100 \% polarization of the spin split bands. The measured polarization of "only" 10 \% is mainly due to the large overlap between the lines with opposite spin. \\
Figure 3(a) shows the y-polarization spectra for a 10 ML thick Pb film for $k_{y}=0$ and different $k_{x}$ values. This data can be fitted as described before to obtain the energy separation of the bands (solid lines). From this data it is clear that the spin direction changes sign when passing through $\overline{\Gamma}$ and that at this point the bands are degenerate. In order to illustrate the sensitivity of SARPES a series of simulated polarization spectra is displayed in Figure 3(b) which result from varying the energy spacing of two Gaussians with opposite spin polarization (inset). The change in polarization amplitude within the 1 meV steps is remarkably large, which explains the high accuracy which can effectively be obtained with spin resolved ARPES.\\
Viewed from an external reference frame the spin polarization is first counterclockwise and changes sign with increasing binding energy. Calculated back to the spin quantization axis we first cut through the band with spin up and then through the state with spin down. This means that the direction of the spin quantization axis of state \textit{A} and \textit{B} is as labeled in Figure 2(d), which is opposite to what has been found for the Au(111) surface state\cite{Hoesch1}. As will be explained below, this reversal of spin polarization direction is a direct consequence of the origin of the Rashba splitting in metallic QWS.\\
Figure 3(c) shows a spin polarization spectrum and the resulting spin resolved energy distribution for $k_{y}=-0.08$ \AA $^{-1}$ and $k_{x} = 0$ for a 22 ML thick Pb film. At this coverage two quantum well states are occupied within the silicon band gap, and both of them show Rashba type spin splitting. For the states at 0.15 and 0.4 eV this splitting is 14 and 15 meV respectively. These results indicate that the splitting remains almost constant as a function of coverage, which is a direct result of the unconventional origin of the splitting in the present system.\\
As shown in Figure 3(d) the energy separation of the bands varies between 11 and 15 meV at $\bf{k}$ $\approx 0.1$ \AA$^{-1}$ for coverages between 6 and 22 ML. Based on the observation that the exact binding energy of a QWS depends on the local boundary conditions\cite{Altfeder} it is fair to assume that the intrinsic linewidth of the QWS will not be below 20 meV\cite{comment2}. This means that even with ideal instrumental resolution and without thermal broadening it will not be possible to resolve the two spin split lines in spin integrated ARPES.\\
The Rashba parameter can be determined from half the slope of the energy spacing versus in-plane momentum (Figure 4) and is found to be $\alpha_{R} = 0.04 \pm 0.005$ eV \AA\ . The respective values are 0.33 eV \AA\ and 0.07 eV \AA\ for the Au(111) surface state\cite{Hoesch1} and semiconductor heterostructures\cite{Nitta}. The momentum splitting obtained from this value is $k_{0} = 0.035 \pm 0.002$ \AA$^{-1}$ (0.012 \AA$^{-1}$ and 0.028 \AA$^{-1}$). The Rashba effect in the Pb QWS presented here is thus comparable in size to that in InGaAs/InAlAs heterostructures, with the additional advantage that the former are formed in much thinner structures and are accessible by surface sensitive techniques.\\
\begin{figure}[!htb]
\begin{center}
\includegraphics[width=0.45\textwidth]{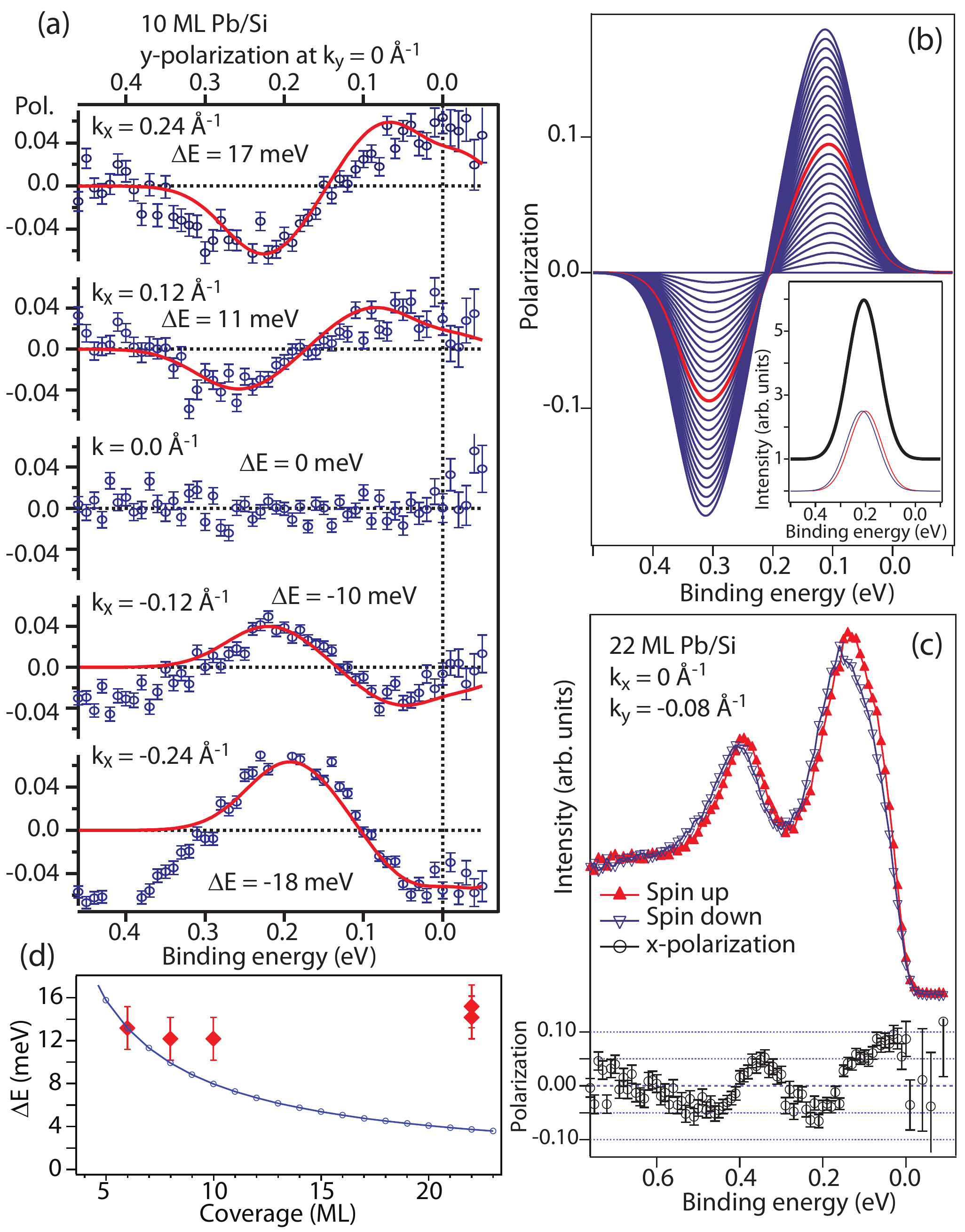} 
\caption{{(color online) (a) Measured (symbols) and fitted (solid line) polarization spectra for a 10 ML thick Pb film at $k_{y} = 0$. (b) Resulting polarization spectra for two Gaussians with opposite spin (lower inset) where the energy spacing is varied in steps of 1 meV. (c) Polarization spectrum and resulting spin resolved spectra for a 22 ML thick Pb film obtained at $k_{y}=-0.08$ \AA $^{-1}$ and $k_{x} = 0$. (d) Measured (red diamonds) spin splitting as function of coverage at $k_{y}=-0.08$ \AA $^{-1}$ and $k_{x} = 0$, the blue circles show the intuitively expected 1/thickness dependence.}}
\label{Fig3}
\end{center}
\end{figure}
Having shown the formation of Rashba type spin-orbit split bands we now turn to the question of what determines the magnitude and sign of the Rashba effect in metallic QWS. As mentioned above, QWS can be regarded as standing electron waves between the metal-vacuum and metal-substrate interface. At each of these interfaces the wave function will experience a phase shift depending on the potential gradient\cite{Chiang}. This phase shift will shift the weight of the wave function away from the atom core position and create a non-zero $\vec{l} \cdot \vec{s}$ product. This means that although the origing of the spin splitting lies at the interface, spin-orbit coupling itself occurs throughout the whole layer. Consequently the size of the splitting can remain constant as a function of thickness, or, depending on the exact wave function, even increase.\\
The influence of the two interfaces is opposite in direction and therefore partly cancels. The spin splitting as measured by SARPES is the resulting net effect and is determined by the difference between the interfaces. The sign of the net Rashba splitting thus depends on whether the phase shift is larger at the metal to vacuum interface or at the metal to substrate interface. If the former is larger the sign will be the same as for the Au(111) surface state. Our data show that the sign is reversed, which indicates that the phase shift is larger at the Pb/Si interface. A change in the electronic or structural properties of this interface will thus alter the Rashba type spin-orbit splitting of the QWS.\\
\begin{figure}[!htb]
\begin{center}
\includegraphics[width=0.45\textwidth]{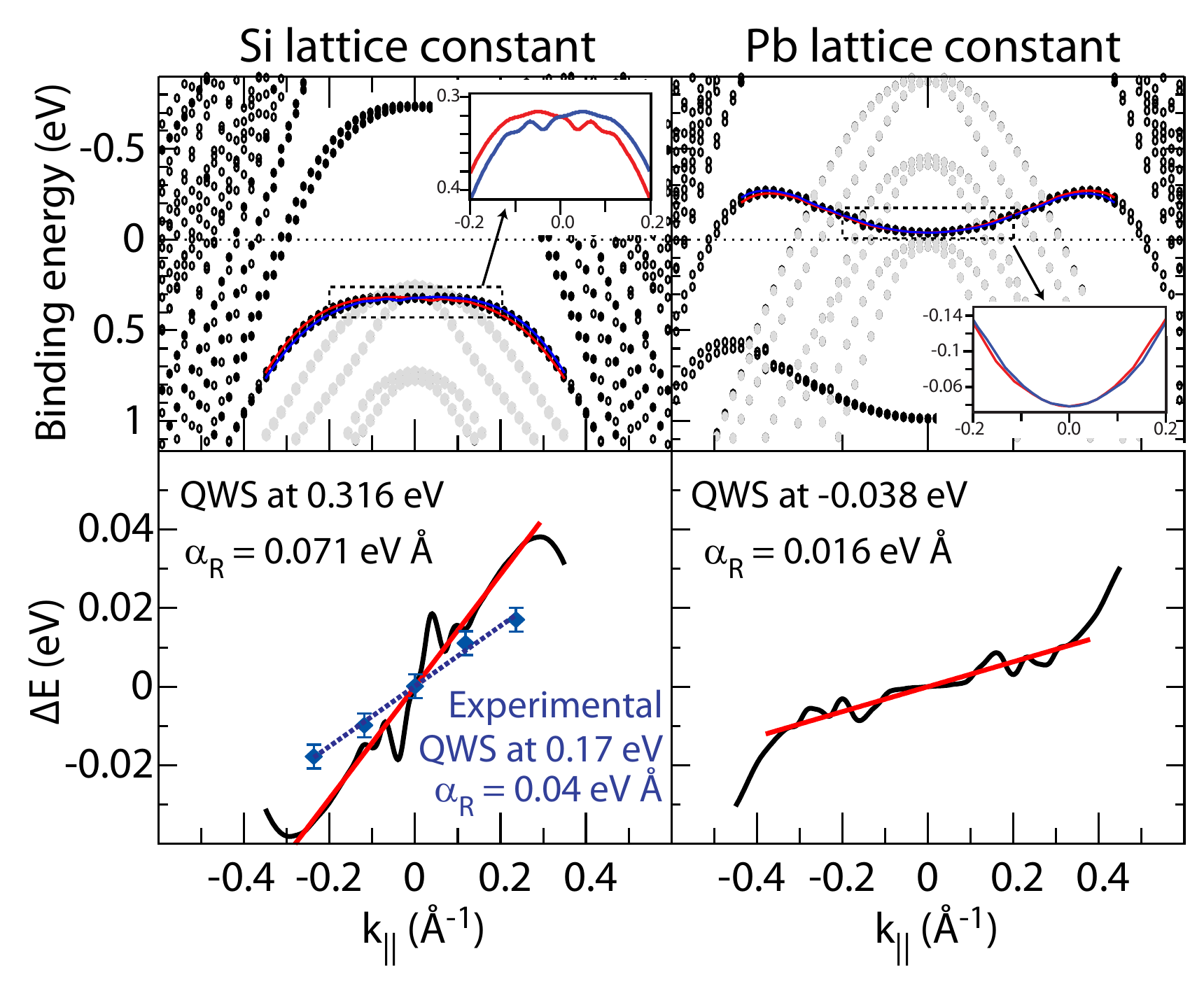}
\caption{{(color online) DFT calculations for 10 ML Pb on Si(111) on the Si (left) and Pb (right) lattice constant. The Si states are shown in gray, Pb states located in the topmost layers in black. The spin-split bands are indicated by colored lines and the size of this splitting is shown in the lower panels together with experimental results. The red lines in the lower panels are fits to obtain the Rashba parameter. The oscillations in energy splitting are the result of hybridization between the QWS and substrate bands in the calculations.}}
\label{Fig4}
\end{center}
\end{figure}
This interpretation is corroborated by density functional theory (DFT) calculations for 10 ML thick Pb films placed on 6 ML of Si\cite{comment3}. The Pb and Si lattices are incommensurate and the correct interface structure can not be included properly. Therefore, the calculations have been performed for two situations; first where the lattice spacing of the Pb overlayer is expanded to match the Si lattice spacing and second where the Si lattice spacing is contracted to match that of the overlayer. The results of these calculations are shown in the left and right panel of Figure 4. From these results it is first of all clear that the high effective mass of the Pb QWS can be explained by assuming a Pb lattice spacing somewhere in between these extremes.\\
The two spin-orbit split bands of the QWS around the centre of the SBZ are nicely reproduced, where the magnitude of the splitting is similar to what is obtained in the SARPES measurements. Furthermore the direction of the spin polarization is also opposite to what is found for the Au(111) surface state. The dependence on the coupling to the substrate suggests that the Rashba effect for QWS in thin metal films can be strongly influenced by interface engineering, including the possibility to reverse the spin polarization direction. A further advantage is that the binding energy of the QWS can be altered by changing the layer thickness or through the selection of the substrate and interface structure. Considering the relatively easily accessible superconductor transition temperature of Pb, this might provide a model system to study the correlation between Rashba splitting and superconductivity\cite{Cappelluti}.\\
In conclusion, we have presented the first observation of Rashba type spin-orbit splitting of quantum well states in ultrathin metal films. The magnitude of the spin splitting found for Pb on Si(111) is too small to be detected by spin integrated ARPES. The magnitude of the splitting, the spin polarization direction and the absence of dependence on the layer thickness are explained by the net effect of competing effects at both interfaces. This makes QWS in thin metal films an ideal candidate for the manipulation of the Rashba effect through interface and surface engineering.
\begin{acknowledgements}
Technical support by M. Kl\"{o}ckner, F. Dubi, and C. Hess is gratefully acknowledged. The measurements have been performed at the Swiss Light Source, Paul Scherrer Institut, Villigen, Switzerland.
\end{acknowledgements}


\begin{thebibliography}{}
\bibitem{comment1}
Because of the inherently low efficiency of Mott detectors the energy and angular resolution were sacrificed in the spin resolved measurements up to 150 meV and 1 degree, respectively.
\bibitem{comment2}
The data in references \cite{Upton} and \cite{Dil2} show a linewidth of about 100 meV, these results should be deconvoluted with the instrumental and thermal broadening to obtain the intrinsic linewidth.
\end{thebibliography}
\end{document}